\documentclass[useAMS,usenatbib]{mn2e}
\usepackage{graphicx}
\usepackage{url}

\title[Mufasa \& LIGO]{Using Galaxy Formation Simulations to optimise LIGO Follow-Up Observations}
\author[Antolini, Caiazzo, Dav\'e \& Heyl]{Elisa Antolini$^{1}$, Ilaria Caiazzo$^{2}$, Romeel Dav\'e$^{3}$, Jeremy S. Heyl$\thanks{Email:
    heyl@phas.ubc.ca; Canada Research Chair}^{2}$ \\
  $^{1}$Dipartimento di Fisica e Geologia, Universit\`a degli Studi di Perugia, I-06123 Perugia, Italia \\
  $^{2}$Department of Physics and Astronomy, University of British
  Columbia, 6224 Agricultural Road, Vancouver, BC V6T 1Z1, Canada\\
  $^{3}$Faculty of Natural Science, University of the Western Cape, Private Bag X17, Bellville 7535, Republic of South Africa\\
}
\begin{document}
\date{Accepted ---. Received ---; in original form ---}

\pagerange{\pageref{firstpage}--\pageref{lastpage}} \pubyear{2016}

\maketitle

\label{firstpage}

\begin{abstract}
  The recent discovery of gravitational radiation from merging black
  holes poses a challenge of how to organize the electromagnetic
  follow-up of gravitational-wave events as well as observed bursts of
  neutrinos.  We propose a technique to select the galaxies that are
  most likely to host the event given some assumptions of whether the
  particular event is associated with recent star formation, low
  metallicity stars or simply proportional to the total stellar mass
  in the galaxy.  We combine data from the 2-MASS Photometric Redshift
  Galaxy Catalogue with results from galaxy formation simulations to
  develop observing strategies that potentially reduce the area of sky
  to search by up to a factor of two relative to an unweighted search
  of galaxies, and a factor twenty to a search over the entire LIGO
  localization region.
\end{abstract}

\section{Introduction}
\label{sec:introduction}
Particular characteristics of individuals are rarely distributed
uniformly over a population.  In fact a small fraction of a
population, even a few percent, can have the majority of a given
attribute.  In this paper we will use this property of the population
of galaxies to optimize electromagnetic follow-up of
gravitational-wave and neutrino transients.

Beginning in September 2015 LIGO started to detect gravitational wave
events from the local Universe \citep{PhysRevLett.116.061102}.  Even
after the addition of the Virgo detector in the Spring 2017, the
localization of many initial candidate events on the sky is coarse
with the ninety-percent confidence regions covering
hundreds or even thousands of square degrees
\citep{2014ApJ...789L...5K,2014ApJ...795..105S,2015ApJ...804..114B,2016LRR....19....1A}.
Some events will have much better localisations on the order of tens
of square degrees.  Either way developing an observing strategy for
follow-up is crucial.  To understand the host environment, the
evolution of the progenitor and to provide tests of cosmology by
yielding an independent measurement of the redshift of the source
requires an electromagnetic counterpart.  On the other hand, what
these electromagnetic counterparts should look like and how long they
should last are uncertain.  Although many have considered the
electromagnetic transients associated with the mergers of binaries
that include a neutron star
\citep[e.g.][]{2016PhRvD..93b4011E,2016arXiv160107711K,2016arXiv160100017D,
  2015arXiv151205435F,2015ApJ...814L..20M,2015PhRvD..92d4028K,
  2015arXiv150807939S,2015arXiv150807911S}, the first discovered
gravitational wave event (GW150914) was almost certainly the merger of
binary black holes.  In this case there are only a few models
\citep[e.g.][]{2015PhRvL.115n1102G,2015MNRAS.452.3419M,2016MNRAS.457..939C,2016ApJ...817..183Y}
that hypothesize the appearance and duration of the electromagnetic
counterparts.  Rapid electromagnetic follow-up of a large portion of
the probable region increases the chance of success in finding a
potential counterpart, and furthermore, it also increases the
likelihood that a potential counterpart indeed accompanied the event.
Over the span of days or weeks, many electromagnetic transients
typically occur, and with the wide variety of models it will be
difficult to associate unambiguously a particular electromagnetic
event with a candidate gravitational-wave event.

Here we will build upon the strategy that \citet{2016MNRAS.462.1085A}
pioneered to use the The Two Micron All Sky Survey extended source
catalogue \citep[2MASS
  XSC,][]{2000AJ....119.2498J,2006AJ....131.1163S} and approximate
photometric redshifts from the 2MASS Photometric Redshift (2MPZ)
catalogue \citep{2014ApJS..210....9B}, to build an efficient observing
plan to follow up gravitational wave transients.  We will use the
results of the {\sc Mufasa} cosmological galaxy formation simulations
\citep{2016MNRAS.462.3265D} to find correlations between the
calculated 2MASS photometry in the simulations and the properties of
the simulated galaxies themselves, total stellar mass, total mass of
low-metallicity stars and the current star formation rate.  Using
these correlations we will create weighted sky maps from the 2MPZ
where the values of each pixel are proportional to the total stellar
mass, total mass of low-$Z$ stars and star formation rate lying in the
given direction over a given redshift range.

\section{Calculations}
\label{sec:calculations}

The objective is to know where to look to increase the probability of
finding an electromagnetic counterpart to a gravitational wave event.
The modelling and analysis of the gravitational waveform as measured
from the LIGO and other sites yields the probability that the observed
waveform {\bf data} from the event resulted from a source at a given
{\bf position}, $P(\mathrm{data}|\mathrm{position})$
\citep[e.g.][]{2016PhRvD..93b4013S}, and this information can be
combined with a galaxy redshift survey to determine where to look
\citep[e.g][]{2014ApJ...784....8H,2015ApJ...801L...1B,2016arXiv160307333S}.
Here the position includes the location of the source on the sky and
in redshift that can be estimated from the gravitational wave
detection.  \citet{2016MNRAS.462.1085A} created full sky maps of the
galaxy density to estimate $P(\mathrm{position})$ and to construct by
Bayes's theorem
\begin{equation}
  P(\mathrm{position}|\mathrm{data}) = \frac{P(\mathrm{position})
    P(\mathrm{data}|\mathrm{position})}{P(\mathrm{data})}
  \label{eq:0}
\end{equation}
where the optimal strategy is to observe those regions of sky where
the probability of a source position given the data,
$P(\mathrm{position}|\mathrm{data})$, is largest.
\citet{2016MNRAS.462.1085A} were agnostic about which galaxies were
most likely to host the event.  However, in principle one could
associate particular types of gravitational events with particular
properties of galaxies.  The most conservative approach may be to
assume that the merger rate is simply proportional to the total
stellar mass of a galaxy, but one could use information about the type
of event to optimize the search further.  For example,
\citet{2015ApJ...806..263D} argued that the first observed events
would come from the merger of black holes with masses of 30-50 solar
masses whose progenitors were low metallicity stars.  
\citet{2016arXiv160204531B} further elaborated on this picture.  In
this case those galaxies with a large mass of low-$Z$ stars would have
a larger probability of harbouring such a source.  A contrasting point
of view is that the black holes are primordial \citep[e.g.][]{PhysRevLett.117.061101}.
In this case, the events
would not be correlated with galaxies at all.
\citet{2016MNRAS.tmp.1537O} examine in detail how the properties and
evolution of a galaxy affect in the rate of compact object mergers in
the galaxy. On the other hand, the formation of neutron stars in
supernovae could generate bursts of gravitational waves that differ
from in-spirals
\citep[e.g.][]{1996PhRvL..76..352B,2009CQGra..26f3001O,2016ApJ...829L..14K},
so in this case, one would focus on star-forming galaxies for the
follow-up. \citet{2015ApJ...814...58D} argue that the merger rate of
neutron-star binaries correlates with the metal abundance of the
galaxy, so the total stellar mass regardless of metallicity would
provide a better estimate of this rate than the mass of low-$Z$
stars. We connect these properties to the quantities available in
large galaxy surveys, in particular the broad-band fluxes, through the
{\sc Mufasa} simulations of galaxy formation.

The {\sc Mufasa} galaxy formation simulations use state of the art
hydrodynamics, star formation, and feedback modules that well
reproduce the global growth of stellar mass in galaxies
\citep{2016MNRAS.462.3265D}, as well as their star formation rates
(SFRs), gas, and metal properties \citep{2016arXiv161001626D}. Of
particular relevance here is that {\sc Mufasa} matches the observed
distribution of specific SFRs at low redshifts, which is key since
high-sSFR galaxies will be especially good follow-up targets.  To
obtain galaxy spectra they identify galaxies, assume each member star
particle is a single stellar population of a given age and
metallicity, and use Flexible Stellar Population Synthesis
\citep{2009ApJ...699..486C,2010ApJ...712..833C}.  to compute its
spectrum assuming a \citet{2003PASP..115..763C} initial-mass function.
They dust attenuate each stellar spectrum based on the extinction
computed from the metal column density along the line of sight to that
star, assuming a \citet{2000ApJ...533..682C} extinction law.  {\sc
  Mufasa} thus also predicts the distribution of galaxies as a
function of absolute magnitude in the 2MASS bands, star-formation
rate, stellar mass and mass of low-metallicity stars, which we take as
the total mass of star particles in each galaxy with $Z<0.01Z_\odot$,
and also the correlations among these quantities because {\sc Mufasa}
generates an ensemble of nearly 7,000 galaxies in an co-moving volume
of $(50 h^{-1} \mathrm{Mpc})^3$.  The value of the solar metallicity
is $Z_\odot=0.02$.  Our choice of the threshold for the definition of
a low-metallicity star is somewhat
arbitrary. \citet{2015ApJ...814...58D} study the merger rates from
stellar populations with $Z=0.1 Z_\odot$ and contrast these with
$Z=Z_\odot$.  We find that the correlations among the galaxies in {\sc
  Mufasa} are similar when one uses $Z<0.001Z_\odot$, so the particular choice
of the threshold is not important.

\begin{figure}
  \includegraphics[width=\columnwidth]{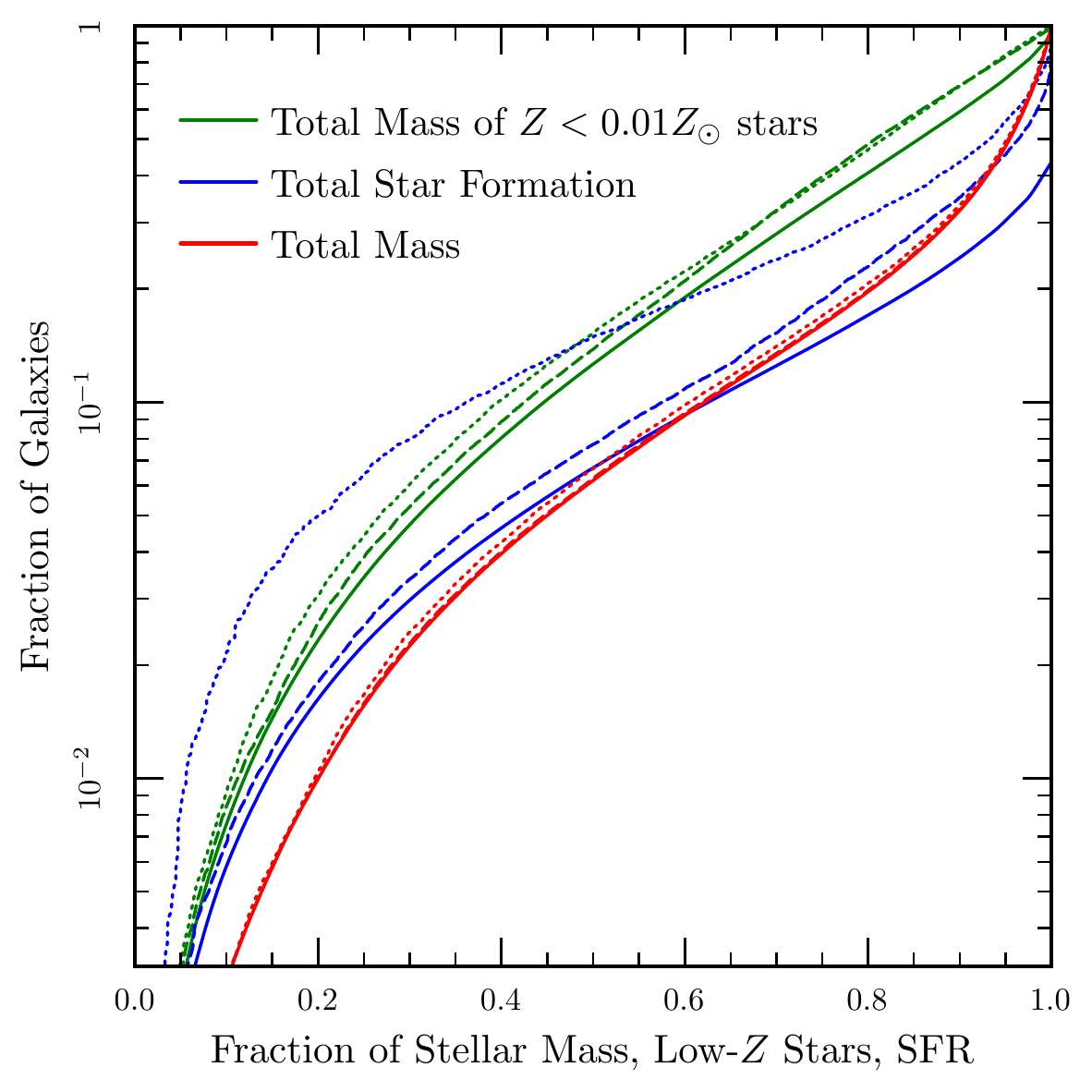}
  \caption{The solid lines traces the minimal fraction of galaxies that
    contain a given fraction of the total stellar mass, low-$Z$
    stellar mass and star formation.  The dotted lines show the
    cumulative fraction if one observes the most luminous galaxies in $K_s$
    first.  The dashed lines use the values estimated from the 2MASS
    luminosities to optimize the observing priority.
  }
  \label{fig:cumgalaxies}
\end{figure}

Fig.~\ref{fig:cumgalaxies} depicts
the fraction of galaxies that contain a given fraction of stellar
mass, low-$Z$ stellar mass and current star formation as the solid
lines.  We immediately can see that half of the stellar mass and half
of the current star formation within the ensemble reside in just about
six percent of the galaxies; therefore, if one can identify which
galaxies have the most star formation for example, one could survey
only six percent of the galaxies to measure half of the young stars in
the local Universe.  This is nearly a factor of ten increase in
efficiency.  The low-metallicity stars in the Universe are spread more
fairly among the galaxies, about one-half of the mass of low-$Z$ stars
resides in about one-eighth of the galaxies.  Here, the increase is
just a factor of four.

The question is whether we can use the observable properties of the
galaxies, in particular their absolute magnitudes in the 2MASS bands
to sort them at least approximately by stellar mass or star formation,
so that one can focus the search to survey the bulk of the stellar
mass or star formation rapidly.  The figure of merit that we will
employ is how much more efficiently we can cover half of the total
star formation or stellar mass than we could achieve by surveying the
galaxies randomly, and how close this efficiency is to the maximum as
depicted in Fig.~\ref{fig:cumgalaxies}.  We fit the logarithm of the
total stellar masses, total low-$Z$ stellar mass, and total star
formation rate of the galaxies within the {\sc Mufasa} simulations as a
linear combination of the 2MASS absolute magnitudes, for example,
\begin{equation}
  \log M_* = A_J M_J + A_H M_H  + A_K M_K + B.
  \label{eq:1}
\end{equation}
The coefficients for the various fits are given in
Tab.~\ref{tab:fitting} along with the Pearson correlation coefficient
$r$.  We have found empirically that the value of $r$ provides a good
estimate of effectiveness of the fit to increase the survey efficiency.
\begin{table}
  \caption{Fitting coefficients for Galaxy Properties}
  \label{tab:fitting}
  \begin{tabular}{l|ccccc}
    \hline
    & $A_J$ & $A_H$ & $A_K$ & $B$ & $r$ \\
    \hline
    Stellar Mass         &  1.87 & -4.07 &  1.79 &  0.75  &  0.98  \\
    Low-$Z$ Stellar Mass & -1.81 &  0.05 &  1.53 &  3.43  & 0.97 \\
    Star Formation       & -3.88 &  6.87 & -3.19 & -2.66  & 0.78 \\
  \end{tabular}
\end{table}
For the stellar mass, the sum of the coefficients ($A_J+A_H+A_K$)
equals $-0.41$, so the total stellar mass is essentially proportional
to the luminosity of the galaxy in the 2MASS bands, a well-known
result and increases most strongly with the increase in the luminosity
in the middle $H-$band.  On the other hand, the sum of the
coefficients for the other two fits are about $-0.2$, indicating that
the star formation rate and mass of low-$Z$ stars within a galaxy are
slower functions of luminosity, perhaps proportional
$L_\mathrm{2MASS}^{1/2}$ or equivalently $M_*^{1/2}$.  The estimator
for low-$Z$ stars favours galaxies that are bluer in the 2MASS bands;
among older stellar populations lower metallicities yield bluer
stars.  Finally, the star formation estimator favours galaxies bright
in the $J$ and $K$ bands, tracing the effects of young stars in the
first case and dust in the second.

In particular the red curves of Fig.~\ref{fig:cumgalaxies} depict the cumulative
stellar mass as a function of the fraction of galaxies survey ordered
by their fitted mass (dashed curve) as well as their luminosities in
the 2MASS bands (dotted curve).  By using the fitted mass we can
survey half of the stellar mass by observing only 6.28\% of the
galaxies; this is only slightly worse that the best possible
performance of 6.20\%.  The fits to the masses using the power-law
relations (Eq.~\ref{eq:1}) from the 2MASS luminosities perform nearly
as well as using the masses themselves.  The solid line is nearly
indistinguishable from the dashed.  One can do nearly as well by just
using the total $K_s$ luminosity as a proxy for the stellar mass as shown by the
dotted lines.  The quality of the fit
for the low-$Z$ stellar masses is slightly poorer
(green curves), but we can still survey half of the low-$Z$
by observing just 14\% of the galaxies, nearly a factor of four
improvement.  In both of these cases, one could achieve nearly as good
performance by simply observing the most luminous galaxies first.
Although one can increase the efficiency of finding the low-$Z$ stars
substantially, the gains are not a large as for the total stellar mass
because the low-$Z$ stars are spread more uniformly among the
galaxies.  If one is undecided whether one wants to survey the entire
stellar population or just the low-$Z$ stars, a reasonable strategy
would simply to look at the most luminous galaxies first; of course,
after observing a given number of galaxies, the completeness of the
survey of the low-$Z$ stars would be worse than that of the total
stellar mass as shown in Fig.~\ref{fig:cumgalaxies}.

The situation is somewhat different for the star formation rate which
does not correlate as well with the 2MASS luminosities as the other
properties do.  The blue curves of Fig.~\ref{fig:cumgalaxies} show that
one can double the efficiency of the search by using the estimated
star formation rate instead of simply observing the brightest galaxies
first.  By using the estimate of the star formation, one could survey
half of the recent star formation in the local Universe by just
observing 7.7\% of the galaxies, a factor of nearly seven improvement
relative to an untargeted search.  If one simply observed the most
luminous galaxies first, one would have to look at 15.8\% of the
galaxies to survey half of the star formation.

\section{Results}
\label{sec:results}
\begin{figure}
  \includegraphics[width=\columnwidth,clip,trim=0 0 0 0.29in]{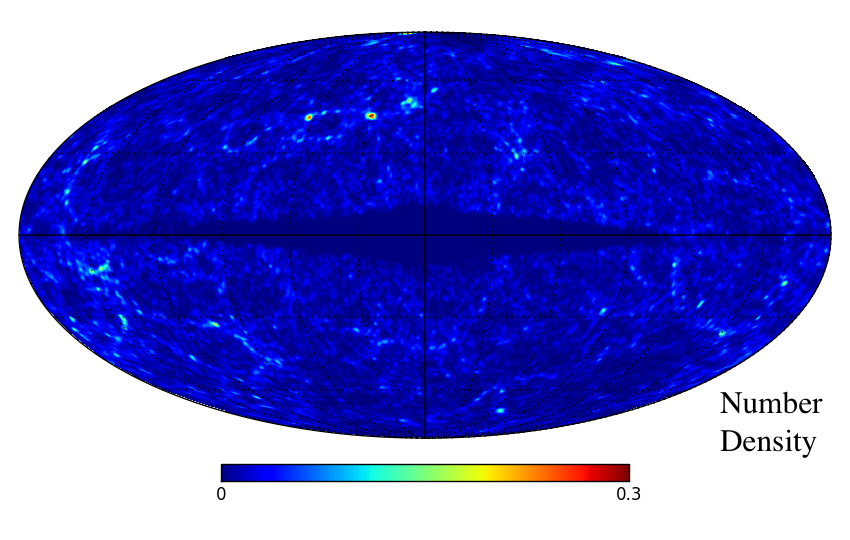}
  \includegraphics[width=\columnwidth,clip,trim=0 0 0 0.29in]{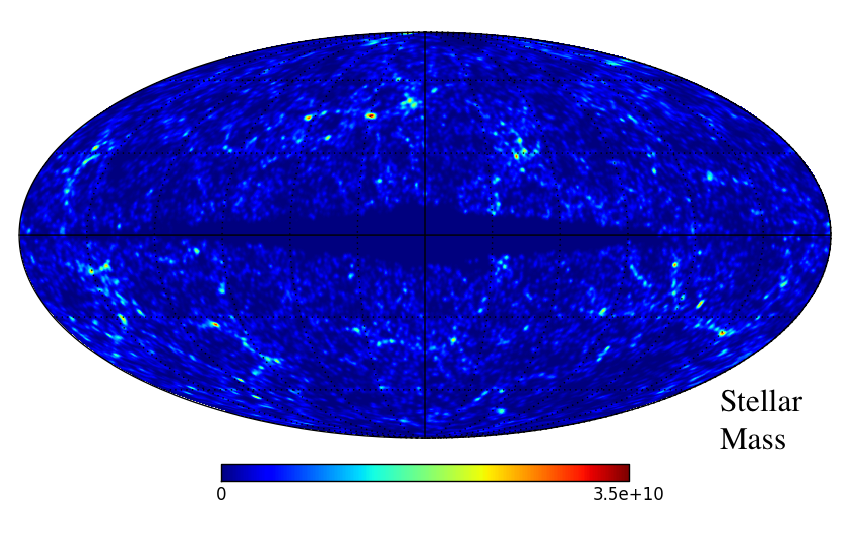}
  \includegraphics[width=\columnwidth,clip,trim=0 0 0 0.29in]{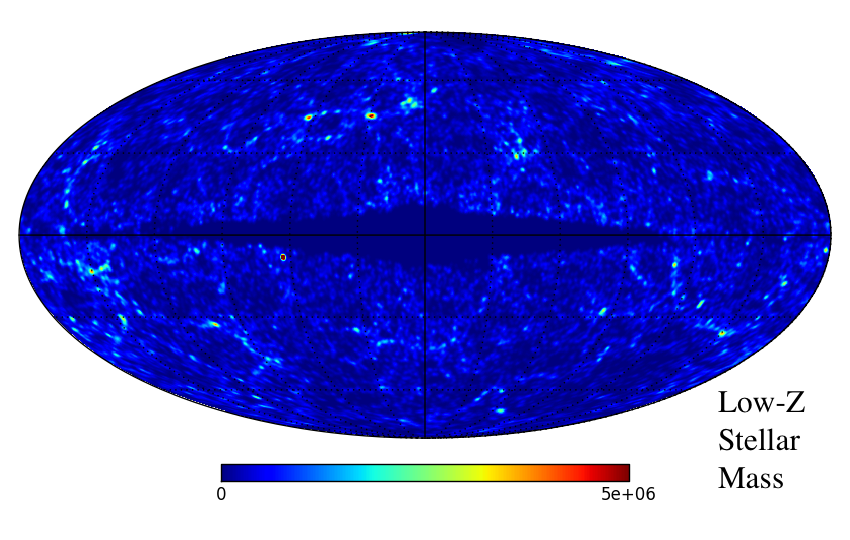}
  \includegraphics[width=\columnwidth,clip,trim=0 0 0 0.29in]{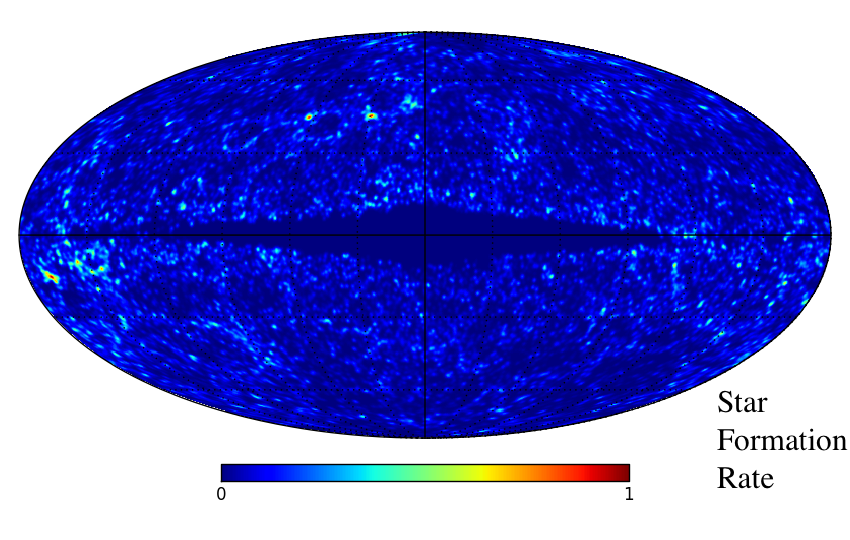}
  \caption{Sky maps of the galaxies with 2MPZ redshifts between 0.03
    and 0.04 weighted by from top to bottom number, total stellar
    mass, total mass of low metallicity stars and current star
    formation rate, smoothed on a scale of 0.6 degrees.  The units are
    number of galaxies, total stellar mass, total low-$Z$ stellar mass
    and total mass of stars formed per year within 0.01 square
    degrees.}
  \label{fig:skymaps}
\end{figure}

To assess the performance of these techniques to create an observing
plan, we will focus on two particular redshift ranges.
\citet{2016MNRAS.462.1085A} examine in detail how the results change
with the telescope field of view and how they would change with the
redshift range.  The first redshift range we will consider is $0.03 <
z < 0.04$.  Fig.~\ref{fig:skymaps} presents the density of galaxies,
stellar mass, low-$Z$ stellar mass and star-formation in the redshift
range using the parameters estimated from the observed 2MASS
luminosities in the 2MPZ, and the fits obtained from the {\sc Mufasa}
simulations.  The upper panel depicts simply the number density of
galaxies on the sky in the 2MPZ \citep[as we examined
  in][]{2016MNRAS.462.1085A}.  The second panel depicts the stellar
mass in the galaxies.  It traces the same structures as in the upper
panel, but the contrast is higher.  Regions of high galaxy density are
relatively stronger peaks in the stellar mass distribution.  The third
panel shows the mass density of low-$Z$ stars.  Here the contrast is
somewhat in between the galaxy density and the stellar mass density,
and furthermore the void regions are less pronounced in the low-$Z$
population.  The lowermost panel depicts the star formation rate.
This again follows the structure in the uppermost panel.  The regions
of the highest star formation correspond to regions of high galaxy
density; however, not every high galaxy density region corresponds to
a high star formation rate.  There is a stochastic element.

\begin{figure}
  \includegraphics[width=\columnwidth]{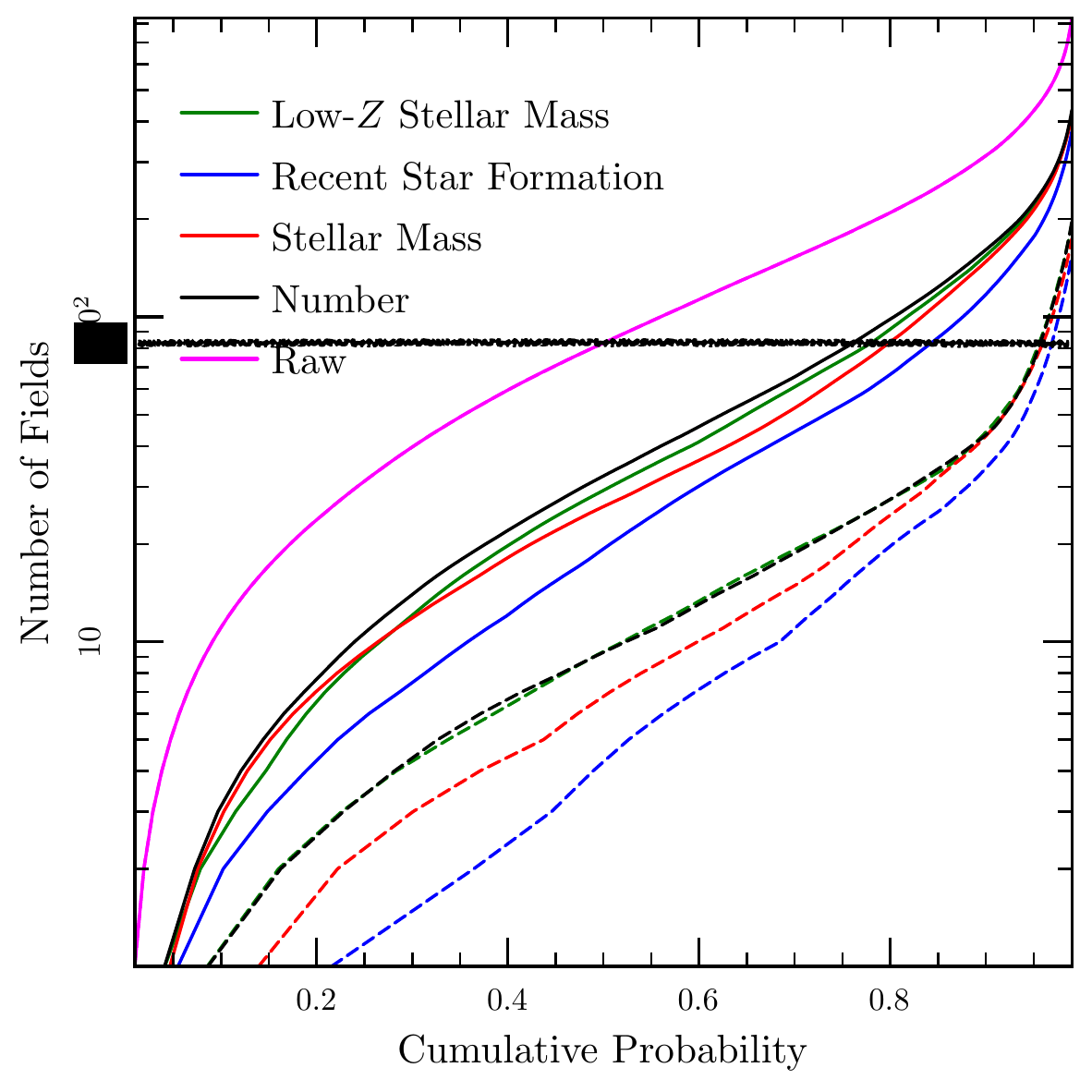}
  \caption{Observing Plans for $0.01 < z < 0.02$ (dashed lines) and
    $0.03 < z < 0.04$ (solid lines) optimized by low-$Z$ stellar mass,
    star formation total stellar mass, galaxy density and the raw LIGO
    probability map.}
  \label{fig:obsplan}
\end{figure}

Now we will examine how using sky maps weighted by galaxy properties
can increase the efficiency of LIGO follow-up.  We choose these
particular redshift ranges for comparision with the previous results
of \citep{2016MNRAS.462.1085A}.  The solid curves in
Fig.~\ref{fig:obsplan} show the results for a redshift range of $0.03
< z < 0.04$.  The constraints from the gravitational-wave detections
are likely to be broader.  We use the Bayesian probability region
calculated by the BAYESTAR algorithm \citep {2016PhRvD..93b4013S} from
\citet{2014ApJ...795..105S} for a LIGO-only detection, that is, before
Virgo is operational. For simplicity, we focus a single field of view
that corresponds to NSIDE = 64 for the HEALPIX map (about one degree
across).  In this redshift range, this corresponds to about 3~Mpc, so
even through the graviational-wave event may be displaced from the
galaxy \citep[e.g.][]{2006ApJ...648.1110B} by up to 1~Mpc, it is still
likely to lie in the same HEALPIX region.  This is the same test as in
\citet{2016MNRAS.462.1085A}.  The improvement by using a galaxy map is
substantial.  However, the additional gains by using the galaxy
properties are modest about 50\% except for the star formation rate
where using the colours of the galaxies in the 2MPZ one could improve
search efficiency by a factor of two.

\begin{figure}
  \includegraphics[width=\columnwidth,clip,trim=0 1.4in 0 0]{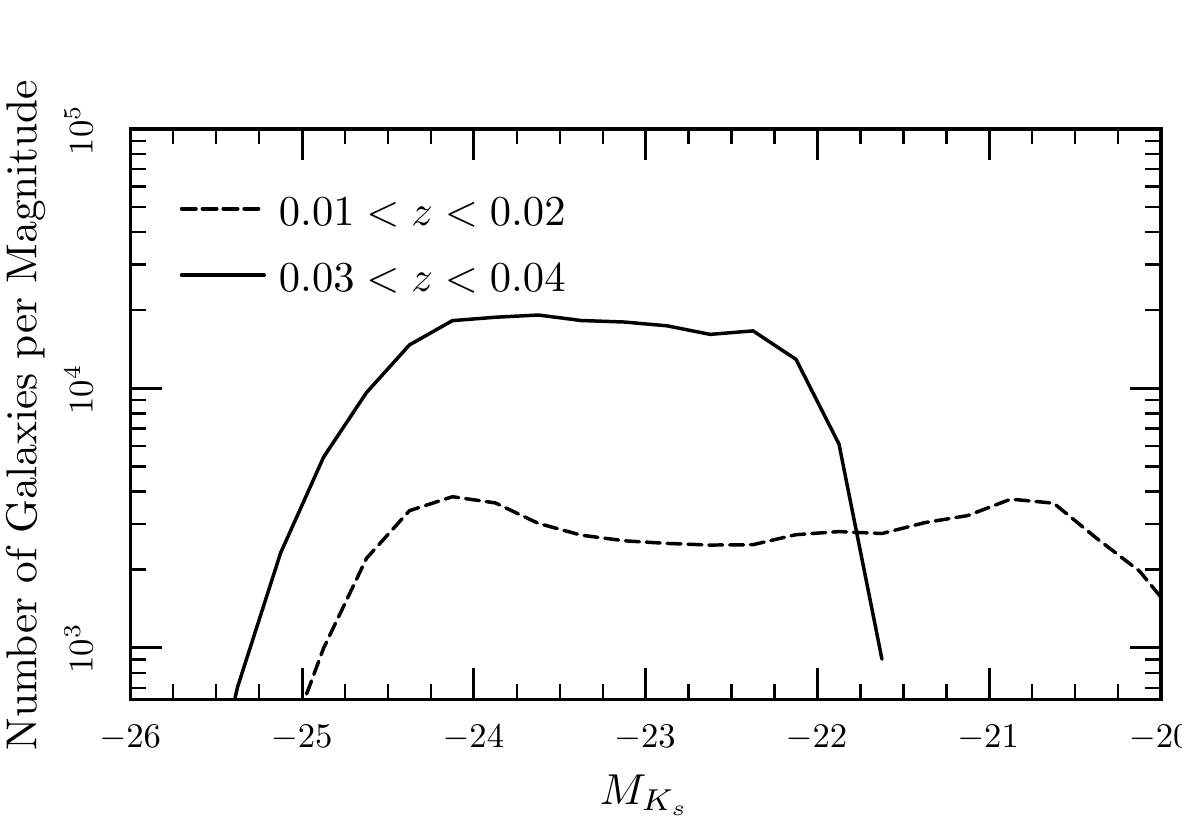}
  \caption{Number of galaxies per unit absolute 
    magnitude for $0.01 < z < 0.02$ (dashed lines) and $0.03 < z <
    0.04$ in the 2PMZ.}
  \label{fig:galhist}
\end{figure}

To understand why the gains are more modest in the 2MPZ survey than in
as simulated in \S~\ref{sec:calculations}, we examine the distribution
of galaxies as a function of absolute $K_s$ magnitude in
Fig.~\ref{fig:galhist} for the galaxies with $0.03 < z < 0.04$ in the
2PMZ.  The luminosities of the galaxies in the catalogue only span a
modest range of about two to three magnitudes.  The magnitude-limited
survey only probes the most luminous galaxies in this redshift range.
These galaxies dominate both the total stellar mass and the mass of
low-$Z$ stars.  On the other hand, the gains for the star formation
rate are more substantial because not all luminous galaxies have
ongoing star formation, so the 2MASS colours can help select those
galaxies where we expect to find star formation and perform a more
efficient search.  Fig.~\ref{fig:galhist} also depicts the
distribution of luminosities in the 2PMZ for a sample of more local
galaxies in $0.01 < z < 0.02$.  In this sample, the range of luminosities
is broader.  The cutoff at high luminosity is at approximately the
same place, but the sample extends to lower luminosities.  In this
case we expect the strategies outlined in \S~\ref{sec:calculations} to
yield stronger gains in efficiency.  These observing plans are
depicted as dashed lines in Fig.~\ref{fig:obsplan}.  To probe the
distribution of low-$Z$ stars, simply following the distribution of
the galaxies themselves yields a similar efficiency to using the
additional information; however, in this small redshift range, there
are only a few nearby galaxies within the LIGO search region and the
2MPZ.  On the other hand, if one is interested in the total stellar
mass or the star formation rate, the gains are larger as expected from
the distribution of galaxy luminosities in the survey.  In fact in
this local sample, one would have to observe only four fields to probe
half of the star formation to compare with ten fields to probe half of
the galaxies and eighty fields to probe half of the LIGO integrated
probability.

\section{Conclusions}
\label{sec:conclusions}

We have presented a technique to optimize the electromagnetic
follow-up observations of gravitational-wave events to focus on the
galaxies that are most likely to host the event.  In principle this
techinque can be combined with further optimizations.  For example
\citet{2016A&A...592A..82G} argue that after the highest probability
regions are identified, adjusting the positions of the pointings
carefully can increase the coverage efficiency by up to 50\% over the
hundred square-degree search region.  Using sky maps weighted by
galaxy properties makes the regions of highest probability much more
structured, so whether an additional 50\% increase in efficiency is
possible is not clear.  Our strategy can also be complemented by
techniques to coordinate multiple telescopes and to account for the
time for the telescope to reach the field or to go from field to field
\citep[e.g.][]{2012arXiv1204.4510S}.

Rapidly searching for the electromagnetic counterparts to
gravitational-wave events is crucial, both because the counterpart may
not last long and because the association of a particular
electromagnetic transient with the gravitational-wave event becomes
less and less significant as time passes.  The initial searches for
electromagnetic counterparts to the GW150914 event discovered many
transients \citep[e.g.][]{2016MNRAS.462.4094S}, but it is unlikely
that any were associated with the source, both because there were not
the expected counterpart and also because the rate of chance
associations is so large over the large search in solid angle and
time.  Strategies as outlined here by reducing the search region
increase the chance of finding a counterpart quickly and also reduce
the false-coincidence rate.

{\noindent \bf Acknowledgements}

The software and galaxy maps used in this paper are available at
\url{http://ubc-astrophysics.github.io}.  We used the VizieR Service,
the NASA ADS service, the SuperCOSMOS Science Archive, the NASA/IPAC
Infrared Science Archive, the HEALPy libraries and arXiv.org. This
work was supported by the Natural Sciences and Engineering Research
Council of Canada, the Canadian Foundation for Innovation, the British
Columbia Knowledge Development Fund and the Bertha and Louis Weinstein
Research Fund at the University of British Columbia.  

\bibliography{mufasa}
\bibliographystyle{apj}

\label{lastpage}
\end{document}